\documentclass[a4paper,twocolumn,11pt]{quantumarticle}
\pdfoutput=1
\usepackage[utf8]{inputenc}
\usepackage[english]{babel}
\usepackage[T1]{fontenc}
\usepackage{amsmath}
\usepackage{hyperref}

\usepackage{tikz}
\usepackage{lipsum}

\usepackage{tkz-graph}
\GraphInit[vstyle = Shade]
\tikzset{
  LabelStyle/.style = { rectangle, rounded corners, draw,
                        minimum width = 2em, fill = yellow!50,
                        text = red, font = \bfseries },
  VertexStyle/.append style = { inner sep=5pt,
                                font = \Large\bfseries},
  EdgeStyle/.append style = {->, bend left} }

\usepackage{slashbox}

\usepackage{lingmacros}

\usepackage{subfigure}
\usepackage{amsfonts}
\usepackage{amssymb}
\usepackage{graphicx,verbatim}
\usepackage{latexsym}
\usepackage{tabularx}
\usepackage{proof}
\usepackage{bussproofs}
\usepackage[toc,page,title,titletoc,header]{appendix}

\usetikzlibrary{decorations.pathmorphing}
\usetikzlibrary{spy,arrows}

\usepackage{rotating}

\usepackage{wasysym}
\usepackage{tikz}
\usetikzlibrary{shapes.misc,automata,positioning,arrows}
\usepgflibrary{shapes.arrows}

\newtheorem{Proof}{Proof}

\newtheorem{Definition}{Definition}
\newtheorem{Theorem}{Theorem}
\newtheorem{Corollary}{Corollary}

\begin{document}

\title[Multi-party Quantum Byzantine Agreement Without Entanglement]{Multi-party Quantum Byzantine Agreement Without Entanglement}

\author{Xin Sun}
\orcid{0000-0001-9320-2522}
\author{Piotr Kulicki} 
\orcid{0000-0001-5413-3886}
\email{kulicki@kul.pl}
\affiliation{Department of the Foundations of Computer Science, the John Paul II Catholic University of Lublin, Poland}
\author{Mirek Sopek}
\affiliation{MakoLab SA, Lodz, Poland}
\maketitle





\begin{abstract}
In this paper we propose a protocol of quantum communication to achieve Byzantine agreement among multiple parties. The striking feature of our proposal in comparison to the existing protocols is that we do not use entanglement to achieve the agreement. There are two stages in our protocol. In the first stage, a list of numbers that satisfies some special properties is distributed to every participant by a group of semi-honest list distributors via quantum secure communication. Then, in the second stage those participants exchange some information to reach agreement. 

\end{abstract}



\section{Introduction}

A fundamental problem in distributed computing is to reach agreement in the presence of faulty processes. For example, a database can be replicated on several computers, which ensures access to the database even if some of the computers are not functional. For the consistency of data, all computers must preserve the same contents. To achieve this goal, a protocol that ensures all computers adopt the same update of the database is needed.
This problem is intuitively formulated as the Byzantine generals problem  \cite{Pease80,Lamport82}:


``Three generals of the Byzantine army want to decide upon a common plan of action:
either to attack (0) or to retreat (1). They can only communicate in pairs by sending messages. One of the generals,
the commanding general, must
decide on a plan of action and communicate it to the other
generals. However, one of the generals might be a traitor, trying to keep the loyal generals from agreeing on a plan. How to find a way
in which all loyal generals follow the same plan?''




If the generals communicate with each other only by pairwise 
classical channels, the Byzantine generals problem is provably unsolvable \cite{Pease80,Lamport82}. Even if pairwise quantum channels are used, it will not help to solve this problem \cite{Fitzi05}. However a
variation of the Byzantine agreement problem, called detectable Byzantine agreement (DBA), can be
solved by using quantum resources. A DBA protocol ensures that, either all loyal generals agree upon a common plan or all abort. In addition, if all generals are loyal, then they agree upon a common plan.

In 2001, Fitzi \textit{et al.} \cite{Fitzi01} presented a DBA protocol for three parties using pairwise quantum channels and entangled qutrits. Cabello \cite{Cabello03} proposed a three-party DBA protocol based on a four-qubit singlet state. 
Iblisdir and Gisin \cite{Iblisdir04} developed an improvement of the protocol of  Fitzi \textit{et al.} \cite{Fitzi01} by 
showing that the DBA problem can be solved by using two  quantum key distribution channels and three classical authenticated channels. Gaertner \textit{et al.} \cite{Gaertner08} introduced a new DBA protocol based on four-qubit entangled state. An experimental implementation
of the protocol is also presented in Gaertner \textit{et al.} \cite{Gaertner08}. A device-independent quantum scheme for the Byzantine generals problem is provided in Rahaman \textit{et al.} \cite{Rahaman15}.

All the aforementioned DBA protocols have only considered
the situation of three parties. In actual distributed computing or blockchains \cite{Nakamoto08,Androulaki18}, the number of parties involved is significantly larger than three. 
Ben-Or and Hassidim \cite{BenOr05}, Tavakoli \textit{et al.} \cite{Tavakoli15} and Luo \textit{et al.} \cite{Luo19}
developed DBA protocols for multiple parties based on high-dimensional entangled states. 
These states are difficult to realize by the current technology. In this paper, we will develop a new DBA protocol for multiple parties. The striking feature of our protocol compared to existing protocols is that no entanglement is used in our protocol. The only quantum technology that we use  is quantum key distribution \cite{Bennett84}, which is a relatively matured topic of research and has recently attracted the interest of the industry.

In Section \ref{A simplified quantum Byzantine agreement} we will introduce our protocol. Then, in Section \ref{Analysis of the protocol}, we  will analyse the properties of our protocol. We shall conclude the paper with future work in Section \ref{Conclusion and future work}.

\section{Quantum Byzantine agreement without entanglement}\label{A simplified quantum Byzantine agreement}

Let us begin with formal definitions of Byzantine agreement.

\begin{Definition}\label{BA}[Byzantine agreement (BA) protocol \cite{Fitzi01}] A protocol among $n$ parties such that one distinct party $S$ (the sender) holds an input value
$x_s \in D$ (for some finite domain $D$) and all other parties (the receivers) eventually decide on an output value in $D$ is said to achieve Byzantine agreement if the protocol guarantees that all honest parties decide on the same output value $y\in D$ and that $y=x_s$ whenever the sender is honest.

\end{Definition}

\begin{Definition}\label{DBA}[Detectable Byzantine agreement (DBA) protocol \cite{Fitzi01}] A protocol among $n$ parties such that one sender $S$ holds an input value
$x_s \in D$ and all other receivers eventually decide on an output value in $D$ is said to achieve detectable Byzantine agreement if the protocol guarantees the following:

\begin{enumerate}
\item Agreement: Either all honest parties abort the protocol, or all honest parties decide on the same output value $y\in D$. 
\item Validity: If all parties are honest, then they decide on the same output value $y= x_s$.

\end{enumerate}

\end{Definition}

Now we introduce our DBA protocol. There are two stages of our protocol. The aim of the first stage is to distribute correlated lists of numbers among the parties involved in the protocol. We will call such a list \emph{reference list} since the parties refer to that lists to check whether the information they receive is trustworthy. Then in the second stage, parties use the reference lists to achieve consensus. We assume the existance of semi-honsest parties to handle the task of reference list distribution. This assumption is similar as in Luo \textit{et al.} \cite{Luo19}. For a party to be semi-honest mens that the party acts according to the description of the protocol, but may 
disclose
information with a certain probability $p$, $0<p<1$. We further assume the parties are connected by pairwise authenticated, error-free, synchronous, classical and quantum channels.

\subsection{Stage 1: List distribution}

\begin{figure}[htb!]
    \centering
    \includegraphics[width=7cm,height=5cm]{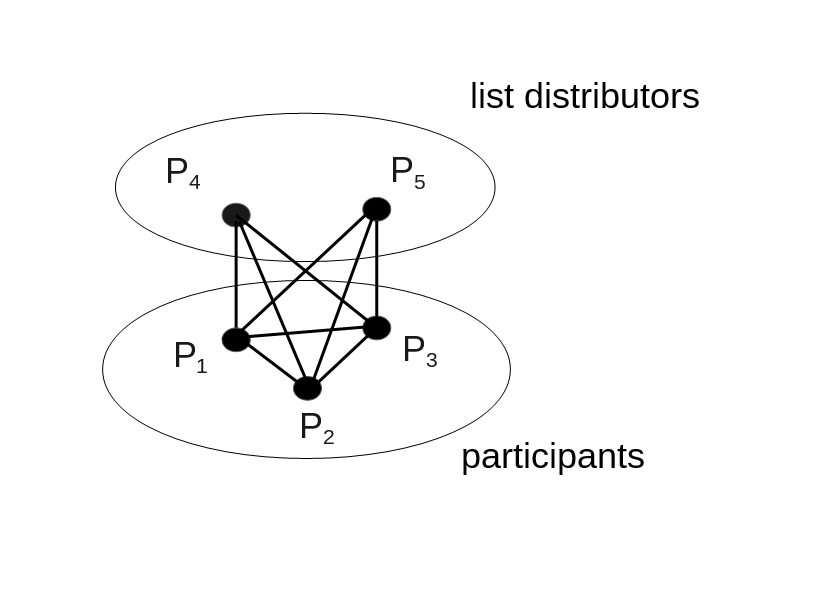}
      \caption{The schema of the system including one sender $P_1$, two receivers $P_2$ and $P_3$, and two list distributors  $P_4$ and $P_5$. All parties are linked via classical and quantum channels (lines in the diagram represent the presence of both channels). In this paper we assume that each participant is connected to any other (except of connections between list distributors which do not exchange any messages between one another), but some of the connections may not necessary need to be both classical and quantum. We plan a detailed study of this issue as  future work. For now we can say that the quantum channels between list distributors and participants are essential.}
    \label{fig:01}
\end{figure}

Let $\{P_1, \ldots , P_n, P_{n+1}, \ldots , P_{n+d}\}$ be a set of parties. Let further $P_1$ be the sender of the DBA protocol, $P_2, \ldots , P_{n}$  be receivers and $P_{n+1}, \ldots , P_{n+d}$ be list distributors. 
To distinguish  the sender and the receivers from the distributors we shall also call the former two  \emph{participants}. 
The schema of the system architecure is presented in Figure \ref{fig:01}.

 We assume that  $P_{n+1}, \ldots , P_{n+d}$ are semi-honest.
For every party $P_i \in \{P_{n+1}, \ldots , P_{n+d}\}$, the task of $P_i$ is to use the technique of quantum secure communication (communicate with the encryption/decryption keys distributed by quantum key distribution) to send a list of numbers $L^i_k$  (a reference list) to each $P_k \in \{P_1, \ldots , P_{n}\}$ such that the following is satisfied:
\begin{enumerate}

\item For all $k\in \{1,\ldots,n\}$, $|L^i_k | =m$ for some integer $m$ which is a multiple of 6.
\item $L^i_1 \in \{0,1,2\}^m$. $\frac{m}{3}$ numbers on $L^i_1$ are $0$.  $\frac{m}{3}$ numbers on $L^i_1$ are $1$.  $\frac{m}{3}$ numbers on $L^i_1$ are $2$.
\item For all $k\in \{2,\ldots,n\}$, $L^i_k \in \{0,1\}^m$.
\item For all $j\in \{1,\ldots,m \}$, if $L^i_1[j]=0$, then $L^i_2[j]=\ldots= L^i_{n}[j] =0$.
\item For all $j\in \{1,\ldots,m \}$, if $L^i_1[j]=1$, then $L^i_2[j]=\ldots= L^i_{n}[j] =1$.

\item For all $j\in \{1,\ldots,m \}$, if $L^i_1[j]=2$, then for all $k\in \{2, \ldots ,n \}$ the probability that $L^i_k[j]=0$ and that $L^i_k[j]=1$ are equal (i.e. the numbers of occurences of $0$ and $1$ are equal in the list).

\end{enumerate}

Distributors create their lists independently so for different $i$ and $j$ the lists $L^i_1$ and  $L^i_1$ may be different (indeed the probability that they are the same is quite small).
%
%
%
%
%
%
%
After the lists are distributed, $P_1, \ldots, P_n$ use sequential composition to form a longer list to be used in the next stage: $L_1= L_{1}^{n+1}\ldots L_{1}^{n+d}, \ldots ,L_n= L_{n}^{n+1}\ldots L_{n}^{n+d} $.
Obviously $L_2 = L_3 = ... = L_n$. We will call the longer lists \emph{combined reference lists}.
Notice that every distributor contributes $\frac{1}{d}$ to the combined reference lists. 

\subsection{Stage 2: reaching agreement}

Now, the parties $P_1,\ldots , P_n$  run the following steps to reach an agreement: 

\begin{enumerate}
\item $P_1$ sends a binary number $b_{1,k}$ to all $P_k$, $k\in \{2,\ldots,n\}$. Together with $b_{1,k}$, $P_1$ sends  to $P_k$ the list of numbers $ID_{1,k}$, which indicate all positions of $b_{1,k}$ on the list  $L_1$. The length of $ID_{1,k}$ is 
to be $\frac{m}{3}$, where $m$ is the length of $L_1$. $P_1$ use $b_{1,k}$ as the final value it outputs.


\item $P_k$ checks the obtained message $(b_{1,k}, ID_{1,k})$ against his own reference list $L_k$. If the analysis of $P_k$ shows that $(b_{1,k},ID_{1,k} ) $ is
consistent with $L_k$, then he 
sends $(b_{1,k},ID_{1,k} ) $ to all other receivers  $P_{j}, j\in \{2,\ldots,n\}$. Here $(b_{1,k},ID_{1,k} ) $ is
consistent with $L_k$ means that for all index $x\in ID_{1,k}$, $L_k[x]=b_{1,k}$. However, if
 $(b_{1,k},ID_{1,k} ) $ is not consistent with $L_k$, 
 then $P_k$ immediately ascertains that $P_1$ is dishonest and
sends  to other receivers $P_{j}, j\in \{2,\ldots,n\}$ message: $\bot$, meaning: ``I have received an inconsistent message''.
To acknowledge the fact that every receiver knows his own output, we formally assume that each of them receives a message from himself.




\item After all messages have been exchanged between the receivers 
every $P_k$ analyzes the data received from $P_2,\ldots , 	P_n$ and acts according to the following criteria:

\begin{enumerate}

\item If there is a set of receivers $H$ with $|H|\geq 2$ such that 
\begin{enumerate}
\item for all $j\in H$, $(b_{j,k}, ID_{j,k})$ is consistent with $L_k$, and 
\item for some $i,j \in H$, $b_{i,k} \neq b_{j,k}$, 
\end{enumerate} 
then $P_k$ sets his output value to be $\bot$.

 \item If there is a set of receivers $H$ with $|H|\geq 2$ such that for all $j\in H$, $(b_{j,k}, ID_{j,k})$ is consistent with $L_k$ and all $b_{j,k} $ are the same, and for all  $i \not\in H$, $(b_{i,k}, ID_{i,k})$ is not consistent with $L_k$, then 
 $P_k$ sets his output value to be $b_{j,k}$.

 \item If there is a set of receivers $H$ with $|H|\geq 2$ such that for all $j\in H$, $(b_{j,k}, ID_{j,k})$ is consistent with $L_k$ and all $b_{j,k} $ are the same, and for all  $i \not\in H$, the message sent by $P_i$ is $\bot$, then $P_k$ sets his output value to be $b_{j,k}$.

\item In all other cases, $P_k$ sets his value to be $\bot$.

\end{enumerate}


\end{enumerate}

The criteria (a) - (d) are crucial for our protocol.
Let us now briefly explain the rationale behind them.  In a nutshell, the most important factor here is the following claim: 

\begin{Theorem}\label{Main theorem}
For all $k, j \in \{2,\ldots, n\}$, $P_k$ believes that $P_j$ is honest whenever $(b_{j,k}, ID_{j,k})$ is consistent with $L_k$.
\end{Theorem}

\begin{Proof}

We prove the theorem by showing that if $P_j$ is dishonest then the probability that $P_j$ sets $(b_{j,k}, ID_{j,k})$ to be  consistent with $L_k$ is extremely small.

Suppose $P_j$ is dishonest. Now $P_j$ wants to send $(b_{j,k}, ID_{j,k})$ to $P_k$ such that  $(b_{j,k}, ID_{j,k})$ is consistent with $L_k$. 
In the case when $P_j$ received a consistent message from $P_1$ it must be the case that $b_{j,k} \neq b_{1,j}$, otherwise $P_j$ would be honest.
Note that in $L_j = L_k$, there are $\frac{m}{2}$ positions on which $b_{j,k}$ appears. But on $L_1$, there are only $\frac{m}{3}$ positions on which $b_{j,k}$ appears. We say that a position $x$ is a \emph{discord} position iff $L_1[x]=2$. If $P_j$ selects a discord position $x$ and puts it into $ID_{j,k}$, then with probability $\frac{1}{2}$ it will be that $L_k[x] \neq b_{j,k}$. To ensure that $(b_{j,k}, ID_{j,k})$ is consistent with $L_k$, $P_j$ has to make a correct choice on all discord positions. The probability of making a correct choice on all discord positions  is $(\frac{1}{2})^{\frac{m}{3}}$, which is extremely small when $m$ is relatively large. Therefore, if it is the case that  
$(b_{j,k}, ID_{j,k})$ is consistent with $L_k$, then $P_k$ can conclude that $P_j$ is honest. 
\end{Proof}

Thus, any receiver $P_k$ can conclusively deduce about any other receiver $P_j$ what follows:
\begin{itemize}
\item If $P_j$ has sent a message consistent with $L_k$, then $P_j$ is honest.
\item If $P_j$ has sent a message inconsistent with $L_k$, then $P_j$ is dishonest.
\item If $P_j$ has sent $\bot$, then $P_j$ may be  honest or dishonest. However, if in this case $P_j$ is honest, then $P_1$ must be dishonest.
\end{itemize} 

The rationale of criterion (a) follows from Theorem \ref{Main theorem}. $P_k$ can conclude that $P_i$ and $P_j$ are honest when $(b_{i,k}, ID_{i,k})$ and $(b_{j,k}, ID_{j,k})$ are consistent with $L_k$. 
Now, if in addition 
$b_{i,k} \neq b_{j,k}$, $P_k$ can safely conclude that the sender ($P_1$) is dishonest. Conequently all the messages are not trustworthy and the output $\bot$ is adequate for the situation. 


As for criterion (b) according to Theorem \ref{Main theorem} we may conclude that all the receivers from the set $H$ are honest and all other are not.  Thus,  $H$ is the set of all honest receivers and their common message is trustworthy.
Criterion (c) is similar to (b).
Receivers from $H$ here are also honest. However in this case some participants who are not in $H$ may also be honest. The honest ones finally will change their output value from $\bot$ to $b_{j,k}$.
For safety reasons with respect to the agreement condition of DBA presented in definition \ref{DBA} by criterion (d) in all other cases honest parties abort our protocol by setting their output to $\bot$.

\section{Analysis of the protocol}\label{Analysis of the protocol}

Now let us analyze  the performance of our protocol under an attack of an adversary. We make the following assumption about the adversary:

\begin{enumerate}
\item The adversary can control a fixed set of participants and let those participants  send arbitrary messages at his will. A participant is dishonest if and only if he is controlled by the adversary. 
The amount of honest participants is  $\geq 3$.
\item The adversary can bribe the list distributors to disclose certain
information. When being bribed, a list distributor will disclose information with probability $p$. 
\item The adversary has unlimited computing power.
\end{enumerate}
In short, the adversary is static, Byzantine and with unlimited computing power.

\begin{Theorem}
Our protocol satisfies agreement and validity under the attack of an adversary. 
\end{Theorem}

\begin{Proof}
It's easy to see that validity is satisfied. Indeed, if none of the participants is controlled by the adversary, then they behave as the protocol specifies. Even if the adversary collects information from a large number of list distributors, the correlated list of numbers will still be correctly distributed. All participants will send consistent messages and the same output value will be established.

We now turn to the proof of agreement. 
First, note that the adversary can hardly have complete information of the combined reference lists  $(L_1,\ldots, L_n)$. By our assumption, every list distributor is semi-honest. They will disclose the content of the list that they distributed with probability $p < 1$, if the adversary bribes them. Since every list distributor contributes only $\frac{1}{d}$ to the lists, to collect complete information about $L_1,\ldots, L_n$, the adversary must bribe all $d$ list distributors and still the probability of collecting complete information is $p^d$, which decreases exponentially as $d$ grows.
For those list distributors that the adversary does not bribe, the adversary cannot collect any information because the lists are distributed by quantum secure communication. The  unlimited computing power the adversary has is not helpful in this case. Therefore, we conclude that the first stage of our protocol can be correctly and safely executed.

Now we consider the second stage.
If the sender 
is honest, then there are at least 2  honest receivers. All honest receivers will receive the same consistent data from the sender. Those honest receivers will forward the same data to other participants. Therefore, according to criterion (a) in our protocol, all honest participants will output the same value as the sender.
If the sender $P_1$ is dishonest, then there are 2 cases:

\begin{enumerate}
\item All honest receivers receive consistent data. In this case there are two sub-cases:

\begin{enumerate}
\item All honest receivers receive the same data. In this case, according to criterion (b), all honest participants will output the same value as the sender.

\item  Not all honest receivers receive the same data. Then, according to criterion (a), all honest receivers will abort the protocol (output $\bot$).

\end{enumerate}

\item  Not all honest receivers receive consistent data. In this case, if there are still two receivers that receive the same and consistent data and all dishonest receivers output $\bot$, then according to criterion (c) all honest receivers will output the same value. Otherwise, according to criteria (a) or (d) all honest receivers will output $\bot$.

\end{enumerate}
 
Therefore, in all possible cases, the agreement is achieved. 
\end{Proof}

The above proof also implies an interesting property of our protocol which is stronger than validity. We present it as a corollary.
\begin{Corollary} Our protocol satisfies the following honest-success property under the attack of an adversary:
if the sender is honest, then all honest parties decide on the same output as the sender.

\end{Corollary}

\section{Conclusion and future work}\label{Conclusion and future work}

We have proposed a protocol of quantum communication to achieve Byzantine agreement among multiple parties. The striking feature of our protocol, compared to existing protocols, is that it does not use entanglement. The success of our protocol relies on the  distribution of sequences of correlated numbers, which in turn relies on the unconditional security of quantum key distribution. 

We also assume the participation  of semi-honest list distributors in the protocol. 
This assumption is the cost to pay for not using entanglement. Since a low-dimensional entanglement can be implemented by  current technology, in the future we will study whether semi-honest distributors could be replaced by a low-dimensional entanglement.
One potential application of our DBA protocol is in the field of quantum blockchain  \cite{Kiktenko17,Sun19,Sun19blockchain}.
In the future we plan to apply our protocol to quantum blockchain to solve particular problems such as auction, lottery and multi-party secure computation.

\subsection*{Acknowledgement}
The project is funded by the Minister of Science and Higher Education within the program under the name ``Regional Initiative of Excellence'' in 2019-2022, project number: 028/RID/2018/19, the amount of funding: 11 742 500 PLN.






\begin{thebibliography}{10}

\bibitem{Androulaki18}
Elli Androulaki, Artem Barger, Vita Bortnikov, Christian Cachin, Konstantinos
  Christidis, Angelo~De Caro, David Enyeart, Christopher Ferris, Gennady
  Laventman, Yacov Manevich, Srinivasan Muralidharan, Chet Murthy, Binh Nguyen,
  Manish Sethi, Gari Singh, Keith Smith, Alessandro Sorniotti, Chrysoula
  Stathakopoulou, Marko Vukolic, Sharon~Weed Cocco, and Jason Yellick.
\newblock Hyperledger fabric: a distributed operating system for permissioned
  blockchains.
\newblock In {\em Proceedings of the Thirteenth EuroSys Conference, EuroSys
  2018, Porto, Portugal, April 23-26, 2018}, pages 30:1--30:15, 2018.

\bibitem{BenOr05}
Michael Ben{-}Or and Avinatan Hassidim.
\newblock Fast quantum byzantine agreement.
\newblock In Harold~N. Gabow and Ronald Fagin, editors, {\em Proceedings of the
  37th Annual {ACM} Symposium on Theory of Computing, Baltimore, MD, USA, May
  22-24, 2005}, pages 481--485. {ACM}, 2005.

\bibitem{Bennett84}
Charles Bennetta and GillesBrassard.
\newblock Quantum cryptography: Public key distribution and coin tossing.
\newblock In {\em Proceedings of IEEE International Conference on Computers,
  Systems and Signal Processing}, pages 175--179, 1984.

\bibitem{Cabello03}
Ad{\'{a}}n Cabello.
\newblock Solving the liar detection problem using the four-qubit singlet
  state.
\newblock {\em Physical Review A}, 68(012304), 2003.

\bibitem{Fitzi05}
Matthias Fitzi, Juan~A. Garay, Ueli Maurer, and Rafail Ostrovsky.
\newblock Minimal complete primitives for secure multi-party computation.
\newblock {\em Journal of Cryptology}, 18(1):37--61, Jan 2005.

\bibitem{Fitzi01}
Matthias Fitzi, Nicolas Gisin, and Ueli Maurer.
\newblock Quantum solution to the byzantine agreement problem.
\newblock {\em Physical Review Letters}, 87(217901), 2001.

\bibitem{Gaertner08}
Sascha Gaertner, Mohamed Bourennane, Christian Kurtsiefer, Ad{\'{a}}nn Cabello,
  and Harald Weinfurter.
\newblock Experimental demonstration of a quantum protocol for byzantine
  agreement and liar detection.
\newblock {\em Physical Review Letters}, 100(070504), 2008.

\bibitem{Iblisdir04}
Sofyan Iblisdir and Nicolas Gisin.
\newblock Byzantine agreement with two quantum-key-distribution setups.
\newblock {\em Physical Review A}, 70(034306), 2004.

\bibitem{Kiktenko17}
E~O Kiktenko, N~O Pozhar, M~N Anufriev, A~S Trushechkin, R~R Yunusov, Y~V
  Kurochkin, A~I Lvovsky, and A~K Fedorov.
\newblock Quantum-secured blockchain.
\newblock {\em Quantum Science and Technology}, 3(035004), 2018.

\bibitem{Lamport82}
Leslie Lamport, Robert~E. Shostak, and Marshall~C. Pease.
\newblock The byzantine generals problem.
\newblock {\em {ACM} Trans. Program. Lang. Syst.}, 4(3):382--401, 1982.

\bibitem{Luo19}
Qingbin Luo, Kaiyuan Feng, and Minghui Zheng.
\newblock Quantum multi-valued byzantine agreement based on d dimensional
  entangled states.
\newblock {\em International Journal of Theoretical Physics}, (58):4025–4032,
  2019.

\bibitem{Nakamoto08}
Satoshi Nakamoto.
\newblock Bitcoin: A peer-to-peer electronic cash system.
\newblock https://bitcoin.org/bitcoin.pdf, 2008.

\bibitem{Pease80}
Marshall~C. Pease, Robert~E. Shostak, and Leslie Lamport.
\newblock Reaching agreement in the presence of faults.
\newblock {\em J. {ACM}}, 27(2):228--234, 1980.

\bibitem{Rahaman15}
Ramij Rahaman, Marcin Wie\ifmmode~\acute{s}\else \'{s}\fi{}niak, and Marek
  \ifmmode~\dot{Z}\else \.{Z}\fi{}ukowski.
\newblock Quantum byzantine agreement via hardy correlations and entanglement
  swapping.
\newblock {\em Physical Review A}, 92, 2015.

\bibitem{Sun19blockchain}
Xin Sun, Mirek Sopek, Quanlong Wang, and Piotr Kulicki.
\newblock Towards quantum-secured permissioned blockchain: Signature,
  consensus, and logic.
\newblock {\em Entropy}, 21(9):887, 2019.

\bibitem{Sun19}
Xin Sun, Quanlong Wang, Piotr Kulicki, and Mirek Sopek.
\newblock A simple voting protocol on quantum blockchain.
\newblock {\em International Journal of Theoretical Physics}, 58(1):275--281,
  Jan 2019.

\bibitem{Tavakoli15}
Armin Tavakoli, Ad{\'{a}}n Cabello, Marek \.{Z}{}ukowski, and Mohamed
  Bourennane.
\newblock Quantum clock synchronization with a single qudit.
\newblock {\em Scientific Reports}, 5(7982), 2015.

\end{thebibliography}



\end{document}